\documentclass[10pt]{IEEEtran}
\usepackage{graphicx}
\usepackage{subfigure}
\usepackage{float}
\usepackage{amsmath}
\usepackage{epstopdf}
\usepackage{cases}
\usepackage{color}
\usepackage{algorithm}
\usepackage{booktabs}
\usepackage{algorithmic}
\usepackage{amsmath}
\usepackage{verbatim}
\usepackage{cite}

\UseRawInputEncoding

\usepackage[textsize=small]{todonotes}
\usepackage{geometry}
\usepackage{lipsum}


\ifCLASSINFOpdf

\else

\fi

\usepackage{caption}
\usepackage{mathtools}

\begin{document}

\title{SLAM for Multiple Extended Targets \\ using 5G Signal}

\author{
	Wangjun Jiang,~\IEEEmembership{Student Member,~IEEE,}
	Zhiqing Wei,~\IEEEmembership{ Member,~IEEE,}
	Zhiyong Feng,~\IEEEmembership{Senior Member,~IEEE,} \\
		Key Laboratory of Universal Wireless Communications, Ministry of Education \\
		Beijing University of Posts and Telecommunications, Beijing, P.R.China, 100876.\\
		Email:  \{jiangwangjun, weizhiqing, fengzy\}@bupt.edu.cn. \\
		{ {Corresponding author: Zhiyong Feng and Zhiqing Wei.}
	}
}

\maketitle

\begin{abstract}

5th Generation (5G) mobile communication systems operating at around 28 GHz have the potential to be applied to simultaneous localization and mapping (SLAM).
Most existing 5G SLAM studies estimate environment as many point targets, instead of extended targets.
In this paper, we focus on the performance analysis of 5G SLAM for multiple extended targets.
To evaluate the mapping performance of multiple extended targets, a new mapping error metric, named extended targets generalized optimal sub-pattern assignment (ET-GOPSA), is proposed in this paper.
Compared with the existing metrics, ET-GOPSA not only considers the accuracy error of target estimation, the cost of missing detection, the cost of false detection, but also the cost of matching the estimated point with the extended target.
To evaluate the performance of 5G signal in SLAM, we analyze and simulate the mapping error of 5G signal sensing by ET-GOPSA.
Simulation results show that, under the condition of SNR = 10 dB, 5G signal sensing can barely meet to meet the requirements of SLAM for multiple extended targets with the carrier frequency of 28 GHz, the bandwidth of 1.23 GHz, and the antenna size of 32.

\end{abstract}


\IEEEpeerreviewmaketitle

\section{Introduction}

\begin{figure*}[h]
	\centering
	\setlength{\abovecaptionskip}{0 mm}
	\subfigure[\scriptsize{\small Real system}.]{\includegraphics[width=0.45\textwidth]{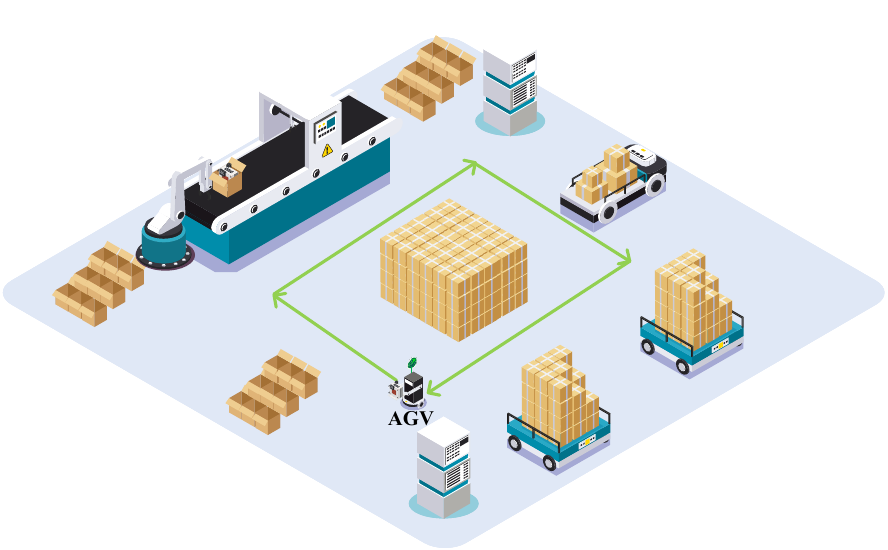}%
		\label{fig:SLAM_system_model_1}}
	\hfil
	\setlength{\abovecaptionskip}{0 mm}
	\subfigure[\scriptsize{\small System model}.]{\includegraphics[width=0.35\textwidth]{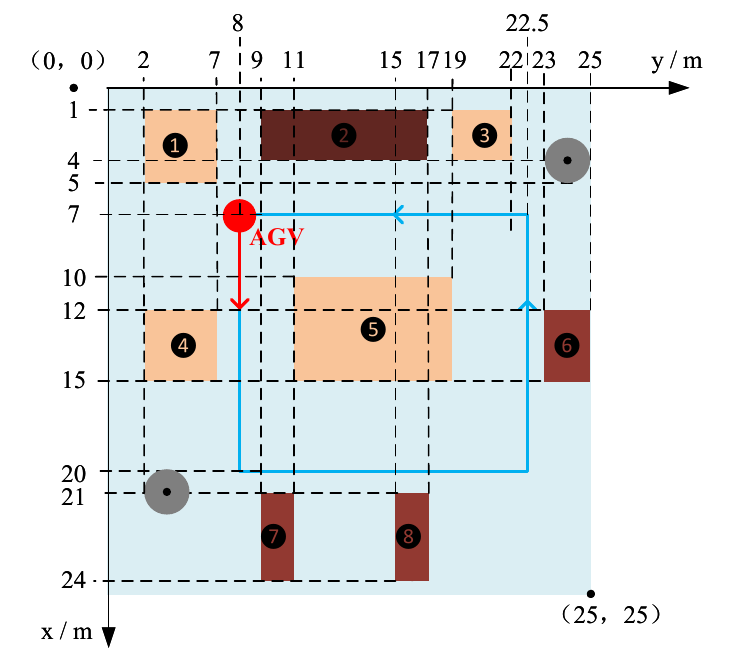}%
		\label{fig:SLAM_system_model_2}}
	\hfil
	\caption{\small SLAM in the industrial flexible manufacturing scenario.}
	\label{fig:SLAM_system_model}
\end{figure*}

In the 5th Generation (5G) mobile communication system, large bandwidth signal and large-scale antenna array bring high resolution in both time-delay and angle domains \cite{[5G_SLAM_1]}, which has led to intense researches in 5G positioning by using both range and angle measurements \cite{[3GPP]}.
This makes it possible for 5G signal sensing to perform simultaneous localization and mapping (SLAM).

Mapping is a crucial component of SLAM, which is challenging due to the complexity of the environment and the need to process high-dimensional data. There are two types of maps that are commonly used in SLAM: feature-based map and occupancy grid map (OGM).
The feature-based map is a sparse map which only has value on the position which has features.
This makes a feature map reliable for localization, but not reliable for navigation and path-planning \cite{[2_5D_Map]}.
On the other hand, OGMs divide the environment into a grid of cells and represent each cell as either occupied or unoccupied by obstacles.
OGMs can be adopted in environments where the obstacles are more diffuse or continuous \cite{[Map_1]}.
In this paper, we focus on the industrial flexible manufacturing scenario with multiple diffuse obstacles, which is suitable for OGM.

To assess and compare the mapping performance of OGM, the similarity between the truth and the estimated set needs to be obtained.
Traditionally, mapping performance assessment relies on intuitive concepts such as the localization error of properly detected targets and the cost of missed targets and false targets \cite{[MTT]}.
The optimal sub-pattern assignment (OSPA) metric is firstly introduced to measure the similarity between distributions of point processes \cite{[SE_GOPSA_11],[SE_GOPSA_12]}.
The application of OSPA for optimal multiple target estimation with known number of targets has been considered in \cite{[SE_GOPSA_14]}.
With unknown number of targets, the application of unnormalized OSPA (UOSPA) for multi-target estimation is proposed in \cite{[SE_GOPSA_19]}.
The generalized OSPA (GOSPA) metric generalizes the UOSPA metric by adding a parameter $\alpha$ to adjust the cardinality mismatch penalty.
In the case of $\alpha = 2$, GOSPA is decomposed into localization errors for properly detected targets, costs for missed targets and costs for false targets \cite{[SE_GOPSA_21],[SE_GOPSA_24]}.

However, the above-mentioned metrics are suitable for point targets, not for extended targets.
To evaluate the mapping performance of multiple extended targets, a new mapping error metric, named extended targets generalized optimal sub-pattern assignment (ET-GOPSA), is proposed in this paper.
Compared with the existing metrics, ET-GOPSA not only considers the accuracy error of target estimation, the cost of missing detection, the cost of false detection, but also the cost of matching the estimated point with the extended target.
Based on the ET-GOPSA, we simulate and evaluate SLAM under different ranging and angular accuracy, further to evaluate the performance of 5G signal sensing for SLAM.

The remaining parts of this paper are organized as follows.
Section \ref{sec:system_model} describes the SLAM system model for the industrial flexible manufacturing scenario.
The design of 5G SLAM based on OGM is introduced in Section \ref{sec:5G_SLAM}.
The performance of SLAM based on 5G signal sensing is analyzed and simulated in section \ref{sec:Simulation}. Section \ref{sec:Conclusion} concludes the paper.

\section{System Model and Signal Model}\label{sec:system_model}

\subsection{System model}\label{sec:system_model_1}

We consider the SLAM system model for the industrial flexible manufacturing scenario, as shown in $\mathrm{Fig.\ \ref{fig:SLAM_system_model}}$.
For the convenience of description, the SLAM system model is constructed by extracting the extended targets in the industrial scenario, as shown in $\mathrm{Fig.\ \ref{fig:SLAM_system_model_2}}$.
There is an automated guided vehicle (AGV) that loops around to detect environmental extended targets and locate its own path.
Specifically, there are eight rectangular and two circular extended targets whose location information is indicated in $\mathrm{Fig.\ \ref{fig:SLAM_system_model_2}}$.
Equipped with $N_t$ transmitted antennas and $N_r$ received antennas, the AGV can send 5G sensing signal and receive echo signal to realize the sensing of the environment.

\subsection{5G Sensing Signal Model}\label{sec:system_model_2}

The 5G orthogonal frequency division multiplexing (OFDM) signal can be expressed as \cite{[OFDM]}
\begin{equation}\label{equ:OFDM_signal}
	x(t) = \sum\limits_{m = 0}^{M - 1} {\sum\limits_{n = 0}^{N - 1} {s(mN + n)} \cdot e^{j2\pi{(f_c + n \Delta f)}t} } {\mathrm {rect}}(\frac{{t - mT}}{T}),
\end{equation}
where $M$ is the number of OFDM symbols, $N$ is the number of subcarriers, $m$ is the OFDM symbol index, $n$ is the index of subcarrier, $s(mN + n)$ is the complex modulation symbol, ${\mathrm {rect}} (\cdot)$ is the rectangle function, $f_c$ is the carrier frequency, ${\Delta f}$ denotes subcarrier spacing. The duration time of each OFDM symbol $T$ contains the elementary symbol duration ${T_{p}}$ and the guard interval ${T_c}$. If ${T_c}$ is larger than the maximum multipath delay, the inter-symbol interference can be eliminated.

\section{5G Signal Sensing for OGM SLAM}\label{sec:5G_SLAM}

\begin{figure}[ht]
	\includegraphics[scale=0.5]{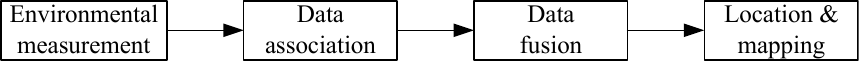}
	\centering
	\setlength{\abovecaptionskip}{2 mm}
	\caption{\small Main modules in SLAM.}
	\label{fig:5G_SLAM}
\end{figure}

In this section, we will introduce a 5G signal sensing based OGM SLAM, as shown in $\mathrm{Fig.\ \ref{fig:5G_SLAM}}$.
Compared with the traditional OGM SLAM scheme \cite{[Lidar_SLAM]}, the SLAM based on 5G signal sensing has great differences in the environmental measurement module, which will be introduced in $\mathrm{Section.\ \ref{sec:5G_SLAM_2}}$.

\subsection{OGM SLAM Modules}\label{sec:5G_SLAM_1}
As $\mathrm{Fig.\ \ref{fig:5G_SLAM}}$ shows, 5G signal sensing based OGM SLAM consists of four main modules: environmental measurement, data association, data fusion, location and mapping.
\begin{itemize}
	\item \textbf{Environmental measurement:}
	The environmental measurement module is a key part of SLAM. The module uses sensor data to estimate the positions of the AGV and the surrounding landmarks.
	Different with visual and lidar SLAM, we adopt the 5G OFDM signal for sensing the surrounding landmarks. Details of OFDM sensing will be introduced in $\mathrm{Section\ \ref{sec:5G_SLAM_2}}$.

	\item	\textbf{Data association:}
	Data association is the process of matching sensor data with map data so that the AGV can determine a map of their location and surroundings.
	The process of data association usually involves two parts: sensing data and mapping data.
	Sensing data is acquired by OFDM sensing, the mapping data is a previously established map of the environment, including obstacles, objects, and ground features.
	In this paper, we select key frames in the sensing data to build the mapping data \cite{[Lidar_SLAM]}.

	\item	\textbf{Data fusion:}
	Data fusion is the process of combining information from multiple sensors to improve the accuracy of positioning and maps for AGVs.
	In this paper, we only analyze the single AGV sensing for SLAM, thus do not consider the data fusion process.

	\item \textbf{Location $\&$ mapping:}
	Location and mapping are the two key tasks of SLAM. Location refers to the process of the AGV to accurately determine its own location in an unknown environment.
	Mapping refers to the process of building environmental models in unknown environments.
	Location requires feature points or topological structure of map, and mapping requires accurate location information. Therefore, the core idea of SLAM is to iterate between location and mapping to improve the accuracy and stability of location and mapping.
	In this paper, we adopt the OGM for the industrial flexible manufacturing scenario with multiple diffuse obstacles.
\end{itemize}
Moreover, to reduce the cumulative error caused by environment sensing, loop detection module is generally adopted in the SLAM system, which can be referred to \cite{[Lidar_SLAM]}.

\subsection{5G signal for Sensing}\label{sec:5G_SLAM_2}
To locate the AGV and map the environment, range and angle parameters are estimated from the received signal by the discrete Fourier transform (DFT) algorithm.

\subsubsection{Range Estimation}\label{sec:5G_SLAM_2_1}
Based on the signal model mentioned in $\mathrm{Section\ \ref{sec:system_model_2}}$, the $n$-th subcarrier $m$-th OFDM symbol of the baseband received signal can be expressed as \cite{[OFDM]}
\begin{equation}\label{equ:OFDM_receive}
	\begin{aligned}
		{\bf Y}(m,n) = \sum_{l=0}^{L_p -1} & {\bf A}(m,n){\bf X}(m,n)\\
		& \cdot e^{- j2\pi n \Delta f \frac{{{2r_l}}}{c_0} } \cdot e^{ \frac{j2\pi {\it{m}}T{{2v_l}{f_c}}}{c_0} }\\
		& + {\bf Z}(m,n),
	\end{aligned}
\end{equation}
where ${\bf A}(m,n)$ is the amplitude of channel information, ${\bf Z}(m,n)$ is the additive gaussian white noise (AWGN), ${\bf X}(m,n)$ is the transmitted modulation signal, $c_0$ is the speed of light, $L_p$ is the number of targets, $r_l$ and $v_l$ are the range and velocity of the $l$-th target, respectively.
The rectangular window function ${\rm rect}(\cdot)$ with time delay can be neglected in baseband processing.
After removing the transmitted information from the received information symbols by an element-wise complex division \cite{[OFDM]}, the elements of division matrix expression can be derived as
\begin{equation}\label{equ:OFDM_signal_1}
	\begin{aligned}
		{({{\bf{S}}_g})_{m,n}} &= \frac{{{\bf Y}} (m,n)}{{{\bf X}} (m,n)} \\
		&=
		{\bf A}(m,n){{\bf k}_r(n)}{{\bf k}_v(m)} + \frac{{{\bf Z}} (m,n)}{{{\bf X}} (m,n)}
	\end{aligned},
\end{equation}
where
\begin{equation}\label{equ:OFDM_signal_2}
	{{\bf k}_r(n)} = \sum_{l=0}^{L_p - 1} e^{- j2\pi {n \Delta f}\frac{{{2r_l}}}{c_0}},
\end{equation}
\begin{equation}\label{equ:OFDM_signal_3}
	{{\bf k}_v(m)} = \sum_{l=0}^{L_p - 1} e^{ \frac{j2\pi {\it{m}}T{{2v_l}{f_c}}}{c_0} } .
\end{equation}
Applying inverse DFT (IDFT) for each column of ${{\bf{S}}_g}$,
\begin{equation}\label{equ:OFDM_signal_33}
	\begin{aligned}
	\mathcal {IDFT} ({{\bf k}_r(n)}) = \frac{1}{N} \sum_{k=0}^{N-1} \sum_{l=0}^{L_p-1} e^{- j2\pi {n \Delta f}\frac{{{2r_l}}}{c_0}} e^{j \frac{2 \pi}{N} ki}, \\
	\quad i = 0,1,...,N,
	\end{aligned}
\end{equation}

the range of target ${r_l}$ can be deduced as follows
\begin{equation}\label{equ:distance}
	r_l \in \left [\frac{{{I^l_{s,m}} \cdot c_0}}{{{2N \cdot \Delta f}}},\frac{{({I^l_{s,m}} + 1) \cdot c_0}}{{{2N \cdot \Delta f}}} \right ),
\end{equation}
where ${I^l_{s,m}}$ is the index of the peak of the IDFT outputs of the ${{m}}$-th column of ${{\bf{S}}_g}$.

\subsubsection{Angle Estimation}\label{sec:5G_SLAM_2_2}
Only two-dimensional positions of industrial scenario are mapped in this paper, so it is assumed that AGVs are equipped with $N_t$ evenly distributed linear arrays.
Thus, the received steer vector of targets can be expressed as \cite{[TVT]}
\begin{equation} \label{equ:AOA_1_1_1}
	\begin{aligned}
		{\bf a}(\Omega_l) &= \sum_{l=0}^{L_p -1}
		{\begin{bmatrix}  1,& e^{j\Omega_l},& \cdots,& e^{j(N_t-1)\Omega_l}  \end{bmatrix} }^T, \in {{\mathcal C}^{N_t \times 1}}
	\end{aligned},
\end{equation}
where $\Omega_l = \frac{2 \pi d}{\lambda} {\mathrm {cos}} (\theta_l)$, $d$ is the distance between two adjacent antenna, $\theta_l$ is the direction of $l$-th target, $\lambda$ is the wavelength of 5G signal.
To estimate $\Omega_l$, we can compute DFT of ${\bf a}(\Omega_l)$
\begin{equation} \label{equ:AOA_1_1_2}
	\begin{aligned}
		\mathcal{DFT}({\bf a}(\Omega_l))
		= \sum_{k=0}^{N_t-1} \sum_{l=0}^{L_p-1} e^{j\Omega_l k} e^{-j \frac{2 \pi}{N_t} ki}, \\ \quad i = 0,1,...,N_t
	\end{aligned}.
\end{equation}
Then, the angle estimated result $\Omega_{l}$ can be derived as
\begin{equation} \label{equ:AOA_1_1_3}
	\begin{aligned}
		\Omega_{l} &\in \left [  \frac{2 \pi I^l_{\theta}}{N_t}, \frac{2 \pi (I^l_{\theta} + 1 )}{N_t}  \right ) \\
		\theta_{l} &\in \left [ {\arccos}(\frac{\lambda I^l_{\theta}}{d N_t }), {\arccos}(\frac{\lambda (I^l_{\theta} + 1 ) }{d N_t } \right )
	\end{aligned},
\end{equation}
where $I^l_{\theta}$ is the index of the peak of $\mathcal{DFT}({\bf a}(\Omega_l)) $.

\begin{figure*}[h]
	\centering
	\setlength{\abovecaptionskip}{0 mm}
	\subfigure[\scriptsize{\small $\Delta R = 0\ m, \Delta \theta = 1^\circ$}.]{\includegraphics[width=0.48\textwidth]{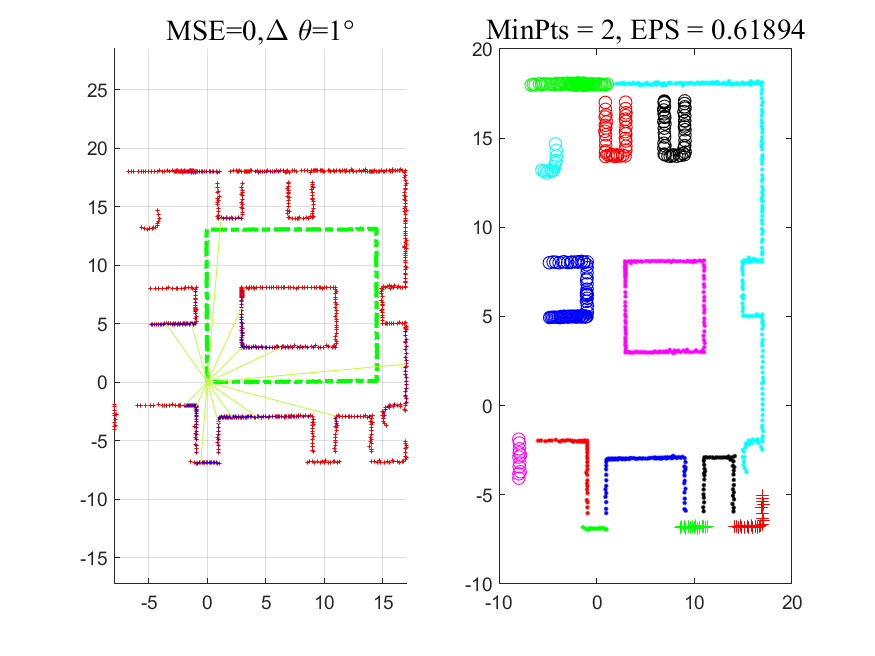}%
		\label{fig:Mapping_1}}
	\hfil
	\setlength{\abovecaptionskip}{0 mm}
	\subfigure[\scriptsize{\small $\Delta R = 0.1\ m, \Delta \theta = 1^\circ$}.]{\includegraphics[width=0.48\textwidth]{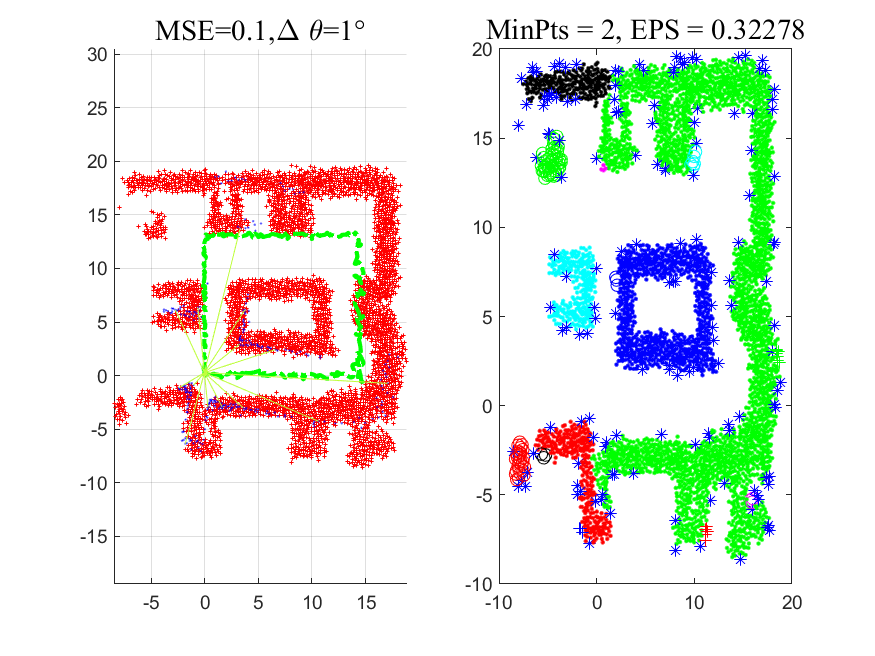}%
		\label{fig:Mapping_2}}
	\hfil
	\setlength{\abovecaptionskip}{0 mm}
	\subfigure[\scriptsize{\small $\Delta R = 0\ m, \Delta \theta = 5^\circ$}.]{\includegraphics[width=0.48\textwidth]{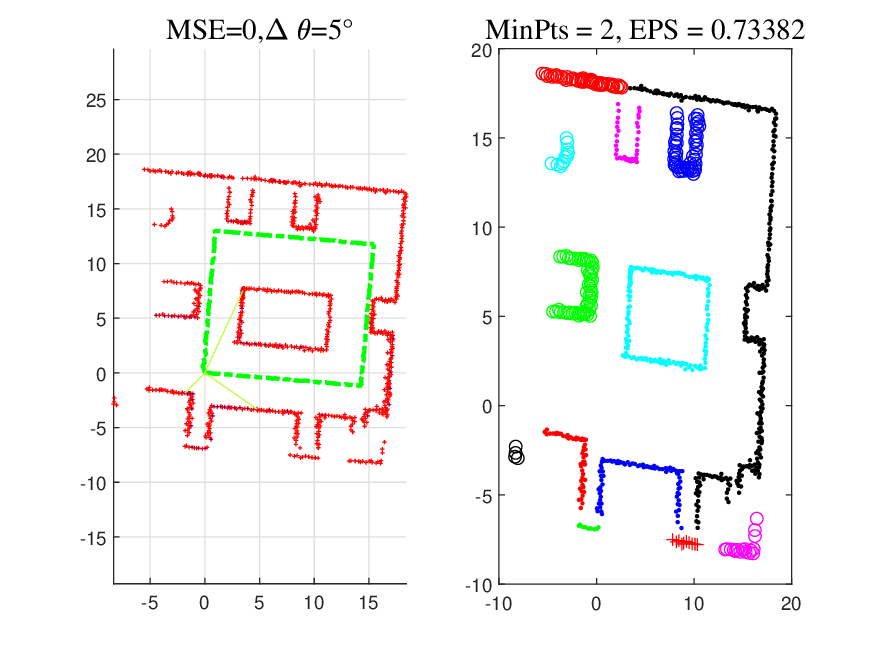}%
		\label{fig:Mapping_3}}
	\hfil
	\setlength{\abovecaptionskip}{0 mm}
	\subfigure[\scriptsize{\small $\Delta R = 0.1\ m, \Delta \theta = 5^\circ$}.]{\includegraphics[width=0.48\textwidth]{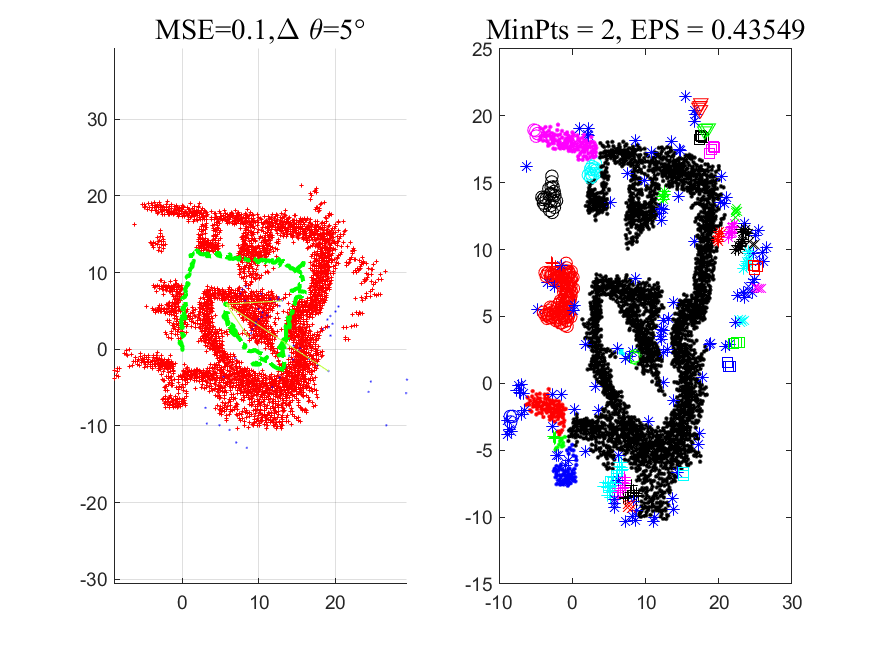}%
		\label{fig:Mapping_4}}
	\hfil
	\caption{\small SLAM with different $\Delta R$ and $\Delta \theta$.}
	\label{fig:Mapping}
\end{figure*}

\section{Performance Analysis and Simulation Results}\label{sec:Simulation}
In this section, we will first simulate the mapping error under different ranging and angular accuracy, and then simulate the mapping error based on 5G signal sensing.
It should be noted that, the SNR adopted in the simulation is denoted the SNR of the received signal, and each simulation in this paper is calculated over 500 Monte Carlo trials.
Simulation parameters used in this section are shown in table \ref{Parameter:simulation} \cite{[5G_signal]}.

\begin{table}[h]
	\caption{Simulation parameters adopted in this paper.}
	\centering
	\label{Parameter:simulation}
	\begin{tabular}{l|l|l}
		\hline
		\hline
		Items & Value & Meaning of the parameter \\ \hline
		$f_c$ & 28 GHz & Carrier frequency \\ \hline
		$\Delta f$ & $120$ kHz & Carrier frequency \\ \hline
		$M$ & 256 & \begin{tabular}[c]{@{}l@{}}Number of OFDM symbols \end{tabular}  \\ \hline
		$N$ & 10240 & \begin{tabular}[c]{@{}l@{}}Number of subcarriers \end{tabular}  \\ \hline
		$T_p$ & 8.3 us & OFDM symbol period \\ \hline
		$T_c$ & 2.08 us & CP period \\ \hline
		$T$ & 10.38 us & The whole OFDM period \\ \hline
		$B$ & 1.23 GHz & Frequency bandwidth \\ \hline
		$N_t$ & 32 & Number of transmitting antenna array \\ \hline
		$N_r$ & 32 & Number of receiving antenna array \\ \hline
		$c$ & 5 & Main parameter in ET-GOPSA \\ \hline
		$p$ & 1 & Main parameter in ET-GOPSA \\ \hline
		$\alpha$ & 2 & Main parameter in ET-GOPSA \\ \hline
	\end{tabular}
\end{table}

\subsection{Mapping error}\label{sec:sim_1}

$\mathrm{Fig.\ \ref{fig:Mapping}}$ shows the location and mapping results under different conditions of the mean squared error (MSE) of ranging $\Delta R$ and the angle estimation error $\Delta \theta$.
As $\mathrm{Fig.\ \ref{fig:Mapping_1}}$ shows, under the condition of $\Delta R = 0\ m, \Delta \theta = 1^\circ$, SLAM has great performance and can resolve most extended targets.
The left figure is the raw location and mapping result by 5G signal sensing, where the green dashed line denotes the location of the AGV path, the red points denote the mapping of the environment, the green solid line denotes the transmitting and receiving signal.
The right figure is the recognition result of extended targets based on the density-based spatial clustering of applications with noise (DBSCAN) method \cite{[DBSCAN]}, different types of points denote different extended targets.
It can be found that under the condition of $\Delta R = 0\ m, \Delta \theta = 1^\circ$, SLAM performs well and can almost distinguish different extended targets.
As $\mathrm{Fig.\ \ref{fig:Mapping_2}}$ shows, under the condition of $\Delta R = 0.1\ m, \Delta \theta = 1^\circ$, SLAM has poor performance and can only resolve partial extended targets. It means that the performance of SLAM is affected by the ranging accuracy.
As $\mathrm{Fig.\ \ref{fig:Mapping_3}}$ shows, under the condition of $\Delta R = 0\ m, \Delta \theta = 5^\circ$, SLAM can almost distinguish different extended targets, but the estimated targets have a certain angle deviation.
Further, when $\Delta R$ is $0.1\ m$ and $\Delta \theta$ is $5^\circ$, SLAM performs poorly and cannot distinguish different extended targets, as shown in $\mathrm{Fig.\ \ref{fig:Mapping_4}}$.

Although, $\mathrm{Fig.\ \ref{fig:Mapping}}$ can qualitatively show the performance of SLAM, it is difficult to measure the mapping error quantitatively. Therefore, ET-GOPSA is presented in this paper to evaluate the mapping performance of multiple extended targets, which will be introduced in $\mathrm{Section\ \ref{sec:sim_2}}$.

\subsection{ET-GOPSA}\label{sec:sim_2}
GOPSA is the typical metric to evaluate the mapping performance of multiple point targets. However, GOPSA cannot evaluate the cost of matching the estimated point with the extended target. Thus, we propose the ET-GOPSA for multiple extended targets, which can be expressed as
\begin{equation} \label{equ:GOPSA_8}
	\begin{aligned}
		& d_{p,\mathcal{E}}^{(c,\alpha)}({\bf \it X}, {\bf \it Y}) = \\
		& \min_{\pi \in {\textstyle \prod_{\left | \bf \it Y \right |}} }
		{\left ( \sum_{i=1}^{\left | \bf \it X \right |}
			\mathcal{E}
			+
			\frac{c^p}{\alpha} (\left | \bf \it Y \right | - \sum_{i=1}^{\left | \bf \it X \right |} |{\bf x}_i|)     \right)}^{1/p} \\
	\end{aligned},
\end{equation}
\begin{equation} \label{equ:GOPSA_9}
	\begin{aligned}
		\mathcal{E} = c + & \min_{k=1,\cdots,|{\bf x}_i|} d^{(c)}(x_{i,k}, y_{\pi (i)})^p  -
		\\ & \min_{k=1,\cdots,|{\bf x}_{-i}|} d^{(c)}(x_{-i,k}, y_{\pi (i)})^p
	\end{aligned},
\end{equation}
\begin{equation} \label{equ:GOPSA_55}
	\begin{aligned}
		d^{(c)}(x, y) = {\rm {min}} (c,||x-y||^2_2)
	\end{aligned},
\end{equation}
\begin{equation} \label{equ:GOPSA_2}
	\begin{aligned}
		\left | \bf \it Y \right | > \left | \bf \it X \right |
	\end{aligned},
\end{equation}
\begin{equation} \label{equ:GOPSA_3}
	\begin{aligned}
		c > 0
	\end{aligned},
\end{equation}
\begin{equation} \label{equ:GOPSA_4}
	\begin{aligned}
		1 \le p \le \infty
	\end{aligned},
\end{equation}
\begin{equation} \label{equ:GOPSA_5}
	\begin{aligned}
		0 < \alpha \le 2
	\end{aligned},
\end{equation}
where $c$ is the maximum allowable localization error, the large the value of the exponent $p$ is, the more the outliers are penalized,
$\bf \it X = \left \{ {\bf x}_1, \cdots, {\bf x}_{\left | \bf \it X \right |}   \right \}$ is the set of truth points. The $i$-th extended target ${\bf x}_i = \left \{ x_{i,1},\cdots,x_{i,|{\bf x}_i|} \right \}$ contains $|{\bf x}_i|$ point targets. ${\bf x}_{-i}$ denotes the extended targets in addition to the $i$-th target,
$\bf \it Y = \left \{ y_1, \cdots, y_{\left | \bf \it Y \right |}   \right \}$ is the set of estimated points, with $\left | \bf \it Y \right |$ being the number of $\bf \it Y$,
${\textstyle \prod_{\left | \bf \it Y \right |}}$ denotes the set of all permutations of $\left \{ 1,2,\cdots,  \left | \bf \it Y \right | \right \}$ and any element ${\pi \in {\textstyle \prod_{\left | \bf \it Y \right |}} } $ denotes a sequence $\left( \pi(1), \pi(2),\cdots,\pi(\left | \bf \it Y \right |)   \right)$.

\begin{figure}[ht]
	\includegraphics[scale=0.6]{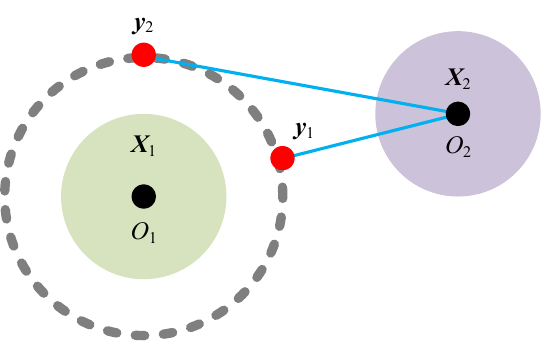}
	\centering
	\setlength{\abovecaptionskip}{2 mm}
	\caption{\small GOPSA $\mathcal{VS}$ ET-GOPSA.}
	\label{fig:ET_GOPSA}
\end{figure}

Compared with GOPSA, ET-GOPSA has the following characteristics
\begin{itemize}
	\item \textbf{Maximize between classes, minimize within classes:}
	The closer the estimated point is to the matched extended target, the smaller ET-GOPSA is; the farther it is from the unmatched extended target, the smaller ET-GOPSA is.
	\item	\textbf{More matching targets:}
	The more estimated points are matched to the extended target, the smaller ET-GOPSA is.
\end{itemize}
$\mathrm{Fig.\ \ref{fig:ET_GOPSA}}$ shows a specific scenario of multiple extended target estimation, where ${\bf x}_1$ and ${\bf x}_2$ are two extended targets, ${y}_1$ and ${y}_2$ are two estimation points. ${y}_1$ and ${y}_2$ have the same distance from ${\bf x}_1$, but different distance from ${\bf x}_2$, then the following conclusions can be drawn
\begin{equation} \label{equ:GOPSA_10}
	\begin{aligned}
		d_{p,\mathcal{E}}^{(c,\alpha)} \left( \left \{{\bf x}_1, {\bf x}_2 \right\}, {y_1} \right) > d_{p,\mathcal{E}}^{(c,\alpha)} \left( \left \{{\bf x}_1, {\bf x}_2 \right\}, {y_2} \right)
	\end{aligned},
\end{equation}
and
\begin{equation} \label{equ:GOPSA_11}
	\begin{aligned}
		d_{p,\mathcal{E}}^{(c,\alpha)} \left( \left \{{\bf x}_1, {\bf x}_2 \right\}, {y_1} \right) > d_{p,\mathcal{E}}^{(c,\alpha)} \left( \left \{{\bf x}_1, {\bf x}_2 \right\}, \left\{ {y_1},{y_2} \right\} \right) \\
		d_{p,\mathcal{E}}^{(c,\alpha)} \left( \left \{{\bf x}_1, {\bf x}_2 \right\}, {y_2} \right) > d_{p,\mathcal{E}}^{(c,\alpha)} \left( \left \{{\bf x}_1, {\bf x}_2 \right\}, \left\{ {y_1},{y_2} \right\} \right) \\
	\end{aligned}.
\end{equation}

\subsection{SLAM Based on 5G Signal Sensing}\label{sec:sim_3}
In this section, we analyze and simulate the performance of SLAM based on 5G signal sensing. The frequency of 5G signal is 28 GHz, the bandwidth is about 1.23 GHz, the number of antenna array is set as 32.
The signal-to-noise ratio (SNR) of 5G signal is set as 10 dB.
For each extended target, four point targets are selected to calculate ET-GOPSA.

\begin{figure}[ht]
	\includegraphics[scale=0.55]{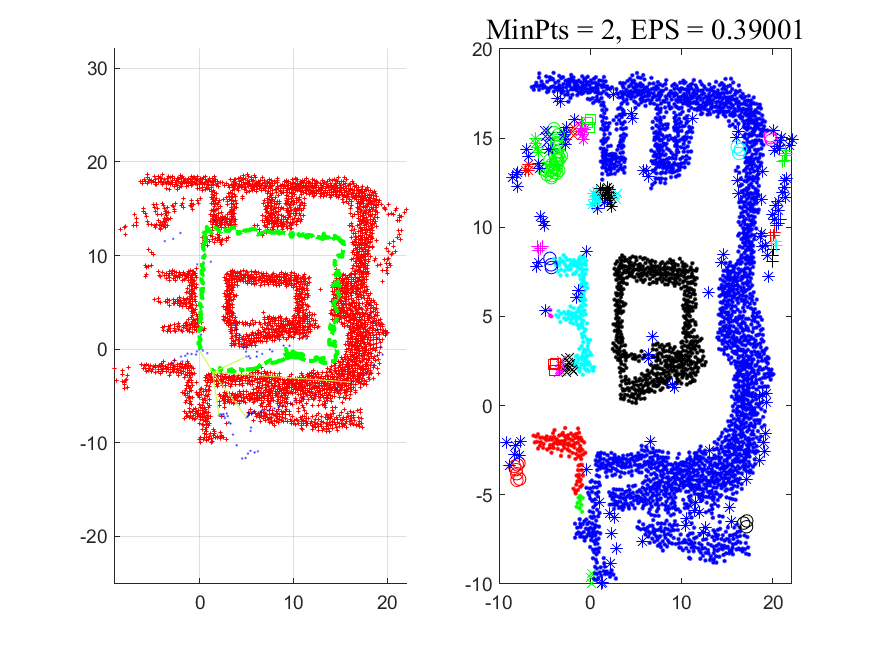}
	\centering
	\setlength{\abovecaptionskip}{0 mm}
	\caption{\small SLAM with 5G signal sensing.}
	\label{fig:5G_SLAM_1}
\end{figure}

\begin{figure}[ht]
	\includegraphics[scale=0.55]{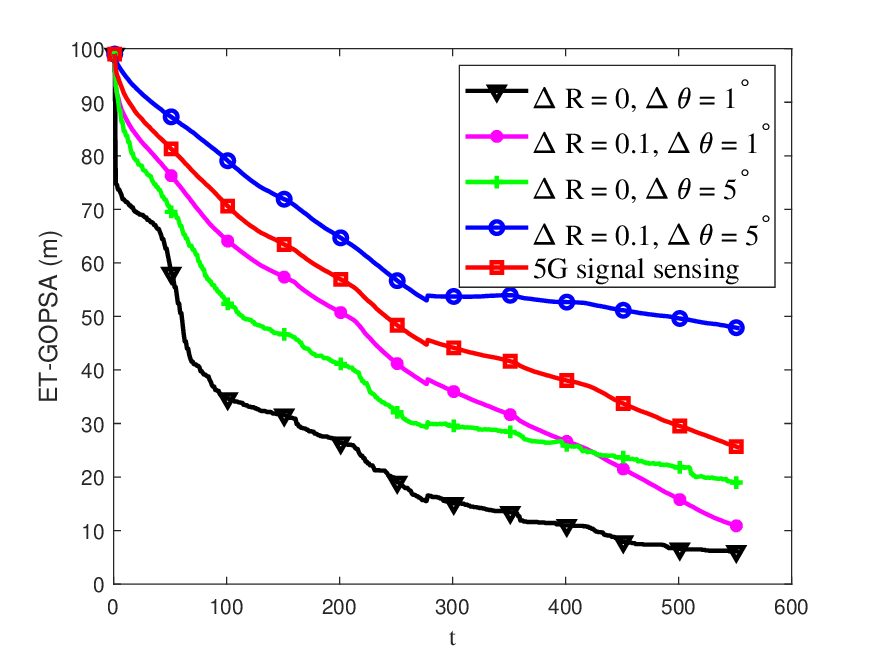}
	\centering
	\setlength{\abovecaptionskip}{0 mm}
	\caption{\small Mapping error of 5G signal sensing.}
	\label{fig:5G_SLAM_2}
\end{figure}

$\mathrm{Fig.\ \ref{fig:5G_SLAM_1}}$ shows the mapping results with 5G signal sensing, where different extended targets are marked in the same color, which means that the extended targets are not clearly detected.
Moreover, the ET-GOPSA of mapping error based on 5G signal sensing is shown in $\mathrm{Fig.\ \ref{fig:5G_SLAM_2}}$. The parameters in ET-GOPSA are set as $c = 10$, $p = 1$, $\alpha = 2$ \cite{[SE_GOPSA_21]}.
As $\mathrm{Fig.\ \ref{fig:5G_SLAM_2}}$ shows, ET-GOPSA is decreasing with the time, which means that more and more environmental targets are sensed.
And ET-GOPSA of 5G signal sensing is larger than that under the condition of $\Delta R = 0.1, \Delta \theta = 1^\circ$ and smaller than that under the condition of $\Delta R = 0.1, \Delta \theta = 5^\circ$.
It further indicates that the single node sensing of 5G signal can barely meet the high accuracy sensing for SLAM.

\begin{figure}[ht]
	\includegraphics[scale=0.55]{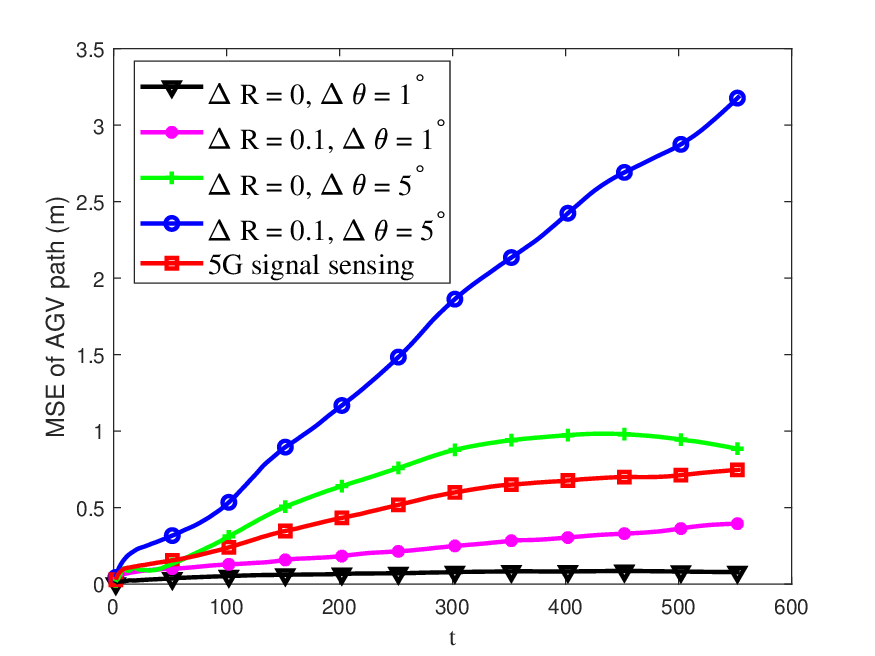}
	\centering
	\setlength{\abovecaptionskip}{0 mm}
	\caption{\small MSE of AGV location.}
	\label{fig:5G_SLAM_3}
\end{figure}

In terms of the location of AGV, MSE of the AGV location is increasing with the sensing time, as shown in $\mathrm{Fig.\ \ref{fig:5G_SLAM_3}}$.
It is because that the error of environment mapping is increasing with the deviation of sensing, which leads to the decreasing positioning accuracy of AGVs based on environmental targets.
MSE of AGV based on 5G signal sensing is larger than that under the condition of $\Delta R = 0.1, \Delta \theta = 1^\circ$ and smaller than that under the condition of $\Delta R = 0.1, \Delta \theta = 5^\circ$. We  can see that the blue line in $\mathrm{Fig.\ \ref{fig:5G_SLAM_2}}$ and $\mathrm{Fig.\ \ref{fig:5G_SLAM_3}}$ exhibit a bad performance of the location of AGV, which denotes that the sensing accuracy of $\Delta R = 0.1, \Delta \theta = 5^\circ$ is hardly sufficient for single-node 5G SLAM.

\section{Conclusion}\label{sec:Conclusion}

In this paper, we focus on the performance analysis of 5G SLAM for multiple extended targets.
To evaluate the mapping performance of multiple extended targets, a new mapping error metric, named ET-GOPSA, is proposed in this paper.
Compared with the existing metrics, ET-GOPSA not only considers the accuracy error of target estimation, the cost of missing detection, the cost of false detection, but also the cost of matching the estimated point with the extended target.
To evaluate the performance of 5G signal sensing in SLAM, we analyze and simulate the mapping error of 5G signal sensing by ET-GOPSA.
Simulation results show that, under the condition of SNR = 10 dB, 5G signal sensing with the carrier frequency is 28 GHz, the bandwidth is 1.23 GHz, and the antenna size of 32 can barely meet to meet the requirements of SLAM for multiple extended targets.
Moreover, compared with the existing metric of mapping error, ET-GOPSA can better evaluate mapping error in the scenario of multiple extended targets.
Through the quantitative analysis of ET-GOPSA, more spectrum resources and antenna resources need to be allocated in order for 5G signal sensing to meet the demands of high-accuracy SLAM for multiple extended targets.
In the future, multi-point cooperative sensing will be adopted to improve the SLAM performance.

\section{Acknowledgment}\label{sec:ack}
This work is supported in part by the National Key Research and Development Program under Grant 2020YFA0711302, and in part by the BUPT Excellent Ph.D. Students Foundation under Grant CX2022207.

\begin{appendices}

\end{appendices}

\bibliographystyle{IEEEtran}
\bibliography{reference}

\ifCLASSOPTIONcaptionsoff
  \newpage
\fi

\end{document}